# Enhancing Motor Imagery Decoding in Brain–Computer Interfaces using Riemann Tangent Space Mapping and Cross Frequency Coupling

Xiong Xiong[1], Li Su[2], Jinguo Huang[1,*], Guixia Kang[1,*]

1. School of Information and Communication Engineering, Beijing University of Posts and Telecommunications, Beijing 100876, China

2. Department of Neuroscience, University of Sheffield, Sheffield S10 2TN, UK

3. School of Automation, Beijing University of Posts and Telecommunications, Beijing 100876, China

* Corresponding author: Jinguo Huang and Guixia Kang

## Abstract

**Objective**: Motor Imagery (MI) serves as a crucial experimental paradigm within the realm of Brain-Computer Interfaces (BCIs), aiming to decoding motor intentions from electroencephalogram (EEG) signals. However, achieving accuracy of MI decoding remains a challenging endeavor due to constraints stemming from data limitations, noise, and non-stationarity. **Method**: Drawing inspiration from Riemannian geometry and Cross-Frequency Coupling (CFC), this paper introduces a novel approach termed "Riemann Tangent Space Mapping using Dichotomous Filter Bank with Convolutional Neural Network" (DFBRTS) to enhance the representation quality and decoding capability pertaining to MI features. DFBRTS first initiates the process by meticulously filtering EEG signals through a Dichotomous Filter Bank, structured in the fashion of a complete binary tree. Subsequently, it employs Riemann Tangent Space Mapping to extract salient EEG signal features within each sub-band. Finally, a lightweight convolutional neural network is employed for further feature extraction and classification, operating under the joint supervision of cross-entropy and center loss. To validate the efficacy, extensive experiments were conducted using DFBRTS on two well-established benchmark datasets: the BCI competition IV 2a (BCIC-IV-2a) dataset and the OpenBMI dataset. The performance of DFBRTS was benchmarked against several state-of-the-art MI decoding methods, alongside other Riemannian geometry-based MI decoding approaches. **Results**: DFBRTS significantly outperforms other MI decoding algorithms on both datasets, achieving a remarkable classification accuracy of 78.16% for four-class and 71.58% for two-class hold-out classification, as compared to the existing benchmarks. **Conclusion**: This paper serves as a testament to the effectiveness and ascendancy of DFBRTS for the realm of MI decoding, highlighting its potential to revolutionize the landscape of more robust MI-BCI applications. **Significance**: The findings is underscored by the effectiveness of DFBRTS in advancing the capabilities of MI-BCI applications, promising enhanced interfaces catering to individuals with motor disabilities.

# Keywords

# 1. Introduction

Motor imagery (MI) is a crucial element in Brain-Computer Interfaces (BCIs), where it enables the translation of imagined movements into actionable commands, facilitating innovative applications in fields like neuro-rehabilitation and human-computer interaction [1]. MI stands as a prevalent manifestation of self-induced mental activity, affording individuals the capacity to mentally simulate movements without physical execution [2]. Decoding MI from EEG signals enables the identification ability of an individual's motor intentions, empowering them to control physical or virtual apparatuses, thereby facilitating human interaction with the environment through BCIs [3]. With the rapid advancements in signal acquisition, signal processing, and machine learning technologies, MI-based systems have exhibited significant potential in providing support to patients with conditions such as stroke [4], spinal cord injuries [5], and amyotrophic lateral sclerosis [6]. Consequently, the accurate MI-EEG decoding would play a crucial role in propelling the advancement of MI-BCI system.

The oscillations observed within the motor cortex in response to MI are commonly referred to as Sensorimotor Rhythms (SMRs) [7]. It is widely acknowledged that various categories of MI are intricately linked to distinct spectro-spatial distributions of SMRs [8]. However, the task of decoding motor intention from MI-EEG signals poses a significant challenge due to factors such as low signal-to-noise ratio, susceptibility to noise interference, and high inter-session variability [9].

Existing researches in the realm of MI-EEG classification, driven by machine intelligence, can be broadly categorized into two primary methods: classical machine learning methods and deep learning-based methods [10]. Classical machine learning methods typically involve the initial extraction of predefined features from MI-EEG data, followed by the utilization of machine learning classifiers such as Linear Discriminant Analysis (LDA) and Support Vector Machines (SVM), for the identification of user's intent. On the other hand, deep learning-based methods often employ an end-to-end neural network architecture to automatically extract features from MI-EEG data and perform classification process [11].

In the realm of classical machine learning techniques for decoding MI-EEG, a pivotal aspect entails the extraction of efficacious features to discern SMRs. Among these features, the most frequently utilized extraction method is the Common Spatial Pattern (CSP) [12]. The CSP algorithm transforms EEG signals by ascertaining an optimal spatial filter that maximizes the variance of one MI task while minimizing that of another MI task. Consequently, the CSP algorithm is well suited for multivariate EEG signal feature extraction [13]. Additionally, diverse enhancement strategies have been integrated in conjunction with the CSP algorithm [12].

To address overfitting concerns associated with the CSP method, several approaches enhance the effectiveness of CSP features through weighting or regularization techniques [14]. Furthermore, to encompass a broader spectrum of features, certain studies consider both sequential dependencies and frequency bands as supplements to CSP [15]. Among these CSP-related methodologies, one of the most prominent research directions is the Subband/Filterbank CSP (FBCSP) [16]. FBCSP acquires CSP features across different frequency bands by introducing filter banks to preprocess EEG signals before CSP feature extraction. Subsequently, an SVM classifier is conventionally employed for classification purposes. Nevertheless, it is noteworthy that these methods have constraints when it comes to extracting spatiotemporal information features from the spectral domain, primarily optimized for binary classification scenarios.

In contrast to the CSP algorithms series, which aim to determine an optimal spatial filter primarily for binary classification, the integration of Riemannian geometry broadens the horizons of MI-EEG decoding beyond spatial filters [17]. Since each dimension of EEG corresponds to different electrodes or channels, coupled with the presence of phase information, the representation of EEG within a multidimensional space exhibits non-Euclidean characteristics. On one hand, Riemannian geometry provides a framework to navigate non-Euclidean spaces, facilitating a more precise characterization of signal relationships [18]. On the other hand, due to its capability to account for the geometric properties on matrix manifolds, Riemannian geometry is better suited for handling the covariance matrices of MI-EEG [19].

One intuitive approach entails the classification of MI-EEG data by computing the Riemannian distance between the covariance matrix and the Riemannian means of different classes within the training data. This approach is commonly known as Minimum Distance Riemannian Mean (MDRM) [18]. Tangent Space Linear Discriminant Analysis (TSLDA) maps the covariance matrices onto the tangent space of the Riemannian manifold, converting them into vectors for feature extraction [18]. Subsequently, dimensionality reduction is performed using Principal Component Analysis (PCA) and Analysis of Variance (ANOVA), with LDA utilized as the classifier. Some advanced methods incorporate supplementary features [20] or extend Riemannian geometry-based features into the temporal and frequency domains to enhance their discriminative capacity [21] [22] [21]. However, the frequency band configurations in these methods typically adhere to conventional approaches, constraining the feature extraction to individual frequency bands and failing to capture interactions between different frequency bands. Moreover, these methods frequently employ traditional dimensionality reduction techniques and machine learning classifiers, which do not adaptively acquire and extract the most pertinent features from the raw data. Consequently, their effectiveness in handling Riemannian geometry-based features is limited.

In recent years, an array of deep learning-based approaches for MI decoding have emerged, showcasing remarkable classification performance. In contrast to traditional, manually engineered feature decoding methodologies, deep learning-based techniques possess the capacity to automatically extract features from EEG signals in a hierarchical manner. For instance, classical deep and shallow Convolutional Neural Networks (CNNs), such as Deep ConvNet and Shallow ConvNet, have been employed for end-to-end EEG data processing [23]. To comprehensively capture spatiotemporal patterns, EEGNet, a compact deep learning framework, made its debut [24]. Employs three convolutional layers, EEGNet proficiently extract spatiotemporal patterns from EEG data. Addressing the varying kernel size requirements among different subjects, the research presented in [25] introduced a mixed-scale CNN. Similarly, the study in [26] conceptualized a multi-branch 3D-CNN, which incorporates a novel 3D representation of EEG data, thus demonstrating the robustness across multiple subjects.

Most recently, a pioneering Filter-Bank Convolutional Network (FBCNet) [27] attained the highest classification accuracy on the BCIC-IV-2a dataset. FBCNet utilizes a filter bank data representation followed by spatial filtering to extract spectro-spatially discriminative features. In contrast to EEGNet, FBCNet omits temporal filters and manually derives frequency-localized signals. These deep learning-based methodologies have significantly bolstered feature extraction capabilities for MI-EEG, often surpassing traditional feature-based MI decoding approaches. However, they do exhibit limitations such as reduced interpretability and limited applicability for datasets with finite samples. Furthermore, deep learning techniques that emphasize frequency bands acquire frequency filters through backpropagation, thereby overlooking the potential advantages of handcrafted spectral bands that could facilitate the unveiling of cross-spectral insights [28].

In addition to the fluctuations in spectral power elicited by different MI tasks, cross-frequency coupling (CFC) may also serve a functional role in MI tasks [29]. CFC entails the mutual influence between oscillatory activities at disparate frequencies, signifying the correlation of neural oscillations across distinct frequency bands [30]. Recent studies have detected characteristic CFC patterns within motor-related cognitive tasks [31]. Moreover, a robust phase-amplitude coupling between α and high γ frequencies has been observed during motor planning, which dissipates during the actual execution of movements [27]. In another investigation, [32] illustrated the effectiveness of motion-related CFC characteristics for a 4-class MI-BCI system, indicating that interactions among multiple frequency bands offer an alternative approach for discriminative representations in MI tasks. Despite the extensive researches on CFC to elucidate neural mechanisms, its integration into MI-EEG decoding has remain somewhat limited [33].

Inspired by pioneering research, this paper introduces a novel MI-EEG decoding approach termed Riemann Tangent Space Mapping using Dichotomous Filter Bank with Convolutional Neural Network (DFBRTS). Specifically, Dichotomous Filter Bank (DFB) and Riemannian tangent space mapping (RTSM) are employed for spatial feature extraction across multiple frequency bands. The incorporation of overlapping and multiscale filter bank settings enhances the efficiency of feature extraction. Furthermore, to further extract and classify the obtained features effectively, a lightweight CNN is appended following the feature extraction stage. Within the CNN architecture, the initial subband-wise convolution is harnessed to extract CFC features, uncovering interactions spanning diverse frequency bands. Subsequently, an RTS-wise convolutional layer extracts information within the RTSM feature space. Additionally, following the approach in [FBMSNet], to mitigate the impact of both between-class and within-class variance, we not only minimize the Cross-Entropy (CE) loss function but also introduce the Center Loss function [34]. The efficacy of DFBRTS is validated on both the BCIC-IV-2a dataset and the OpenBMI dataset. The classification outcomes demonstrate that DFBRTS outperforms baseline methods on both datasets, exhibiting significantly improved classification performance.

The primary contributions of this paper can be summarized as follows:

1) We propose a novel MI-EEG decoding method, DFBRTS, which utilizes Riemannian geometry and the concept of Cross-Frequency Coupling (CFC) for feature extraction, coupled with CNN-based classification. Moreover, under the joint supervision of Cross-Entropy (CE) and Center Loss functions, the learned feature representation by CNN enhances both intra-class compactness and inter-class separability.

2) We introduce a proneering frequency band selection method, christened Dichotomous Filter Bank (DFB), which continuously bifurcates the frequency range of interest to obtain multiscale overlapping frequency band windows. Outcomes from ablation experiments demonstrate the efficacy of DFB in enhancing the method's performance.

3) To the best of our knowledge, we stand as trailblazers in the application of the CFC concept to Riemann Tangent Space (RTS) features within MI-EEG decoding. To extract discriminative CFC features, we engineer a lightweight CNN incorporating subband-wise convolution layers and RTS-features-wise convolution layers.

4) We conduct extensive ablation investigations and numerical experiments, conclusively indicating that DFBRTS outperforms classical and recent methods, including FBCSP, Shallow ConvNet, Deep

ConvNet, EEGNet, and FBCNet. Additionally, it surpasses other RTS-based methods, such as MDRM, TSLDA, CSP-TSM, MTSMS, and FBRTS.

The remainder of this paper is organized as follows. Section II presents an overview of the structure and principles of DFBRTS. Experimental methodology and results are detailed in Sections III and IV, respectively. Section V provides a discussion of the findings, while Section VI offers a concise summary.

## 2. Methods

A. Riemannian Geometry

Riemannian geometry is a broad branch of geometry that holds significant importance in the study of properties and structures of multidimensional spaces in general [35]. It extends beyond the realm of Euclidean geometry by encompassing the geometric properties of curves and surfaces, making geometry applicable in non-Euclidean spaces as well [36]. In Riemannian geometry, the most fundamental concept is that of a Riemannian manifold. A Riemannian manifold is a space that can be described using a Riemannian metric, where each point possesses an associated inner product structure. This Riemannian metric allows us to define geometric concepts on the manifold, enabling us to investigate the properties of non-flat spaces [37]. The Riemannian tangent space, on the other hand, is a linear space in the vicinity of a point on the manifold, representing the tangent plane at that point.

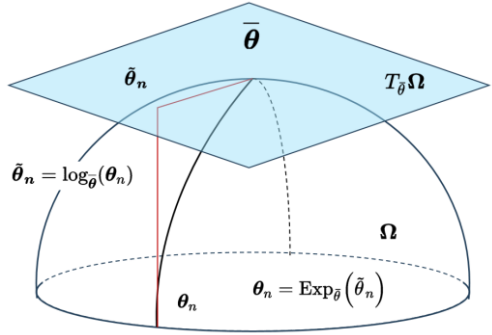

Fig. 1. Corresponding local tangent space at a reference point.

As shown in Fig. 1, for a fixed reference point $\bar{\theta} \in \Omega$, the corresponding tangent space is denoted as $T_{\bar{\theta}}\Omega$. According to the principle of Riemannian geometry, we use logarithmic mapping to project the vector from the Sub-manifold $\Omega$ to its tangent space and use exponential mapping to project the points on the tangent space back to the Sub-manifold, which are as follows:

$$\tilde{\theta}_n = Log_{\bar{\theta}}(\theta_n) = \bar{\theta}^{\frac{1}{2}} \log m \left( \bar{\theta}^{-\frac{1}{2}} \theta_n \bar{\theta}^{-\frac{1}{2}} \right) \bar{\theta}^{\frac{1}{2}} \tag{1}$$

$$\theta_n = Exp_{\bar{\theta}}(\tilde{\theta}_n) = \bar{\theta}^{\frac{1}{2}} \exp m \left( \bar{\theta}^{-\frac{1}{2}} \tilde{\theta}_n \bar{\theta}^{-\frac{1}{2}} \right) \bar{\theta}^{\frac{1}{2}} \tag{2}$$

where $\log m()$ and $\exp m()$ are the logarithmic function and the exponential function of the matrix, respectively.

For MI-EEG signals, the covariance matrix is employed as a representation of spatial information. These covariance matrices are real symmetric positive definite, The Euclidean distance and Riemannian distance between the covariance matrices $M_1$ and $M_2$ are considered as follows:

$$\delta_E(M_1, M_2) = \| M_1 - M_2 \|_F \tag{3}$$

$$\delta_R(M_1, M_2) = \| \log(M_1^{-1} M_2) \|_F = [\sum_{i=1}^{n} \log^2 \lambda_i]^{1/2} \quad (4)$$

where $\delta_E$ is the Euclidean distance and $\delta_R$ is Riemannian distance. $\|.\|_F$ is the Frobenius norm. $\lambda_i$ represents the i-th eigenvalue of $M_1^{-1} M_2$. Which means Euclidean distance represents the shortest distance along a straight-line path, while the Riemannian distance represents the shortest path along with a geodesic search.

Furthermore, considering a collection of covariance matrices $\{M_i\}_{i=1}^{n}$, derived from n trials, we define their arithmetic mean in Euclidean space and their geodesic mean on the Riemannian manifold as follows:

$$\mu(M_1, M_2, \ldots, M_n) = \frac{1}{n} \sum_{i=1}^{n} M_i \quad (5)$$

$$\bar{M} = \vartheta(M_1, M_2, \ldots, M_n) = \arg\min_{M} \sum_{i=1}^{n} \delta_R^2(M, M_i) \quad (6)$$

Where $\mu$ represents the arithmetic mean of $M_i$ in Euclidean space, and, $\bar{M}$ represents the geodesic mean of $M_i$ on the Riemannian manifold. It is worth noting that, there is no closed-form solution for $\bar{M}$. However, the iterative gradient descent algorithm can be used to solve this problem [38].

Furthermore, using the tangent space located at the geometric mean of the whole set trials $\bar{M}$, The covariance matrix $M_i$ of i-th trial can be mapped into a vector using the following equation:

$$V_i = \text{upper}\left(\bar{M}^{-\frac{1}{2}} \log(M_i) \bar{M}^{-\frac{1}{2}}\right) \quad (7)$$

The dimensionality of the vector obtained by applying Riemannian tangent space mapping (RTSM) is $N_c(N_c+1)/2$, where $N_c$ is the row number of the covariance matrix, which is also the number of electrodes in MI-EEG.

## B. DFB

Pre-filtering MI-EEG signals before feature extraction is a widely employed strategy due to the distinct rhythmic responses in brain electrical activity at specific frequency bands [39]. Traditional MI-EEG filtering strategies typically fall into three categories:

1) Fixed-width filters, often targeting frequency ranges like 4-30Hz or 4-80Hz [40-42].
2) Non-overlapping, continuous-width filter banks, typically covering 4-36Hz with 4Hz filter widths, resulting in filter banks such as 4-8Hz, 8-12Hz, and so forth [16, 43, 44].
3) Overlapping, equal-width filter banks that slide with fixed widths and offsets across frequency bands like 4-40Hz or 4-80Hz [22, 45, 46].

However, these filtering approaches have their respective drawbacks. The first category filters only within the region of interest, potentially discarding information from other frequency bands and smaller-sized bands. The second category has fixed-sized segments, which may not be suitable for all subjects. The third category often results in an excessive number of filters, leading to feature redundancy.

Therefore, this paper introduces a binary partition-based EEG filter bank design strategy called Dichotomous Filter Bank (DFB). As illustrated in Figure 2, it begins by selecting a broad region of interest, in the case of MI-EEG, covering 4-80Hz, and includes it as the root filter of the filter bank, referred to as the first-level root filter. Next, the chosen frequency range is recursively divided into two equal sub-frequency ranges, with each added to the filter bank as second-level sub-filters. This process continues iteratively until the *n*-th level is reached, resulting in a filter bank with $2^n - 1$ filters.

Compared to other filtering strategies, DFB offers three key advantages. Firstly, DFB's design approximates the coverage provided by the first two categories of traditional filters, ensuring basic performance. Secondly, DFB is a flexible filter bank design strategy that can adapt to different tasks by choosing different regions of interest and appropriate levels, rather than relying on fixed frequency windows. Lastly, DFB effectively avoids the issues faced by traditional filtering strategies by providing multi-scale, overlapping, and controllable filters. This design, in consideration of the potential role of CFC in MI tasks [29], enables DFB to involve oscillatory activities across a broader range of frequencies, facilitating subsequent CFC feature extraction.

C. CNN and Total Architecture

Table 1. Parameters of CNN

| Layer | Filter | Size | Stride | Activation | Options | Output |
|---|---|---|---|---|---|---|
| Input | | | | | | (1, $2^{level}$-1, $C*(C+1)/2$) |
| Conv2D | 16 | ($2^{level}$-1, 1) | 1 | Linear | Padding=same | (16, 1, $C*(C+1)/2$) |
| BatchNorm | | | | | | (16, 1, $C*(C+1)/2$) |
| Conv2D | 32 | (1, $C*(C+1)/2$) | 1 | LeakyReLu | Padding=same | (32, 1, 1) |
| BatchNorm | | | | | | (32, 1, 1) |
| Reshape | | | | | | (32,) |
| FC | 32 | | | LeakyReLu | | (32,) |
| BatchNorm | | | | | | (32,) |
| FC | Classes | | | Softmax | | (Classes,) |

$C$ represents for number of channels, *level* represents for chosen decomposition level described in Session 2.B.

After applying DFB and RTSM, each MI-EEG trial yields $2^{level} - 1$ features, each of length $N_c(N_c+1)/2$. These features possess significant structured meaning, where the $2^{level} - 1$ dimensions represent different frequency bands, and the $N_c(N_c+1)/2$ dimensions represent the tangent space feature vectors. Given the limited sample size of MI-EEG data, a small Convolutional Neural Network (CNN) with two convolutional layers is employed for further feature extraction and classification. In the first convolutional layer, a subband-level kernel is used to learn relationships between different frequency bands, i.e., CFC features. The second convolutional layer is utilized to capture relationships at the tangent space feature level. Subsequent layers are used for classification. The specific parameters of the CNN network structure are outlined in Table 1. Furthermore, we incorporate the center loss alongside the cross-entropy loss to enhance intra-class compactness while improving inter-class separability. The final loss function is formulated as follows:

$$\mathcal{L}_{\text{total}} = \mathcal{L}_{CE} + \lambda \mathcal{L}_C = -\sum_{k=1}^{classes} y_k \log \hat{y}_k + \frac{\lambda}{2} \sum_{i=1}^{m} \| f_i - c_{y_i} \|_2^2$$

where $y$ is the true label and $\hat{y}$ is the predicted label. classes denotes the number of classes. Where $f_i \in \mathbb{R}^{d \times 1}$ is the feature vector of sample $i$. $c_{y_i} \in \mathbb{R}^{d \times 1}$ is the feature center of the class to which sample $i$ belongs. Following the suggestion in [34] and the execution in [47], we update the feature centers based on the mini-batch in each iteration instead of the entire training set.

A flow chart of the overall approach is shown in Fig. 2.

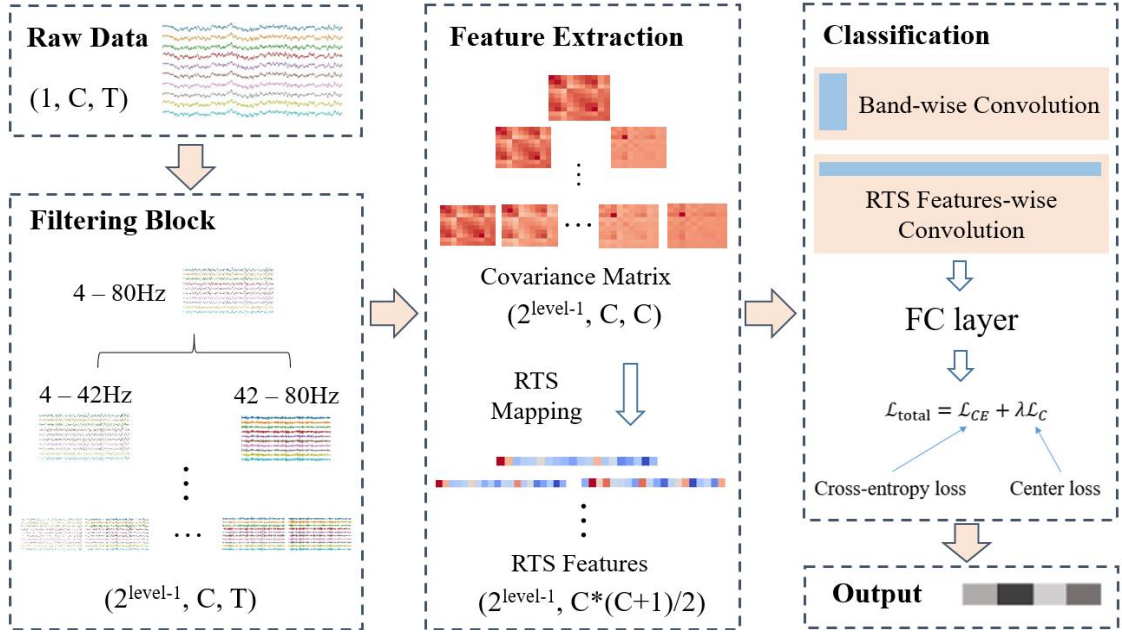

Fig. 2. Flow chart of the overall approach.

## 4) Experiments

A. Datasets

Two datasets were used for evaluating DFBRTS and baseline methods, brief description is provided as follows:

BCIC-IV-2a [48]: The BCIC-IV-2a dataset contains MI-EEG collected from 9 different subjects. Dataset 2a consists of 4 MI tasks (left hand, right hand, feet, and tongue). Each subject in dataset 2a has 2 sessions, and each session has 72 trials across four categories. The EEG data were recorded with 22 electrodes sampled at 250 Hz.

OpenBMI [49]: The OpenBMI dataset is a large benchmark containing 2 sessions of 2-class MI-EEG data (left hand, right hand) from 54 right-handed healthy subjects, among whom 38 subjects were naive BCI users. Each session consists of training and test phases, and each phase has 100 trials with balanced right and left-hand MI tasks.

B. Benchmark Methods

We compare the performance of DFBRTS with ten methods, half of them based on Riemannian geometry while anther half is not. An overview of the compared methods is described as follows:

FBCSP-SVM [16]: FBCSP first uses a set of bandpass filters in conjunction with the CSP algorithm. Filters of FBCSP follows the second filter bank strategy described in Session 2.B. SVM is selected as the classifier. Of note, FBCSP was the best performing method for the BCI Competition IV 2a dataset during the competition.

Deep ConvNet [23]: Deep ConvNet has four convolution max-pooling blocks, with a specialized first block designed to handle the EEG input, followed by three standard convolution-max pooling blocks and a dense softmax classification layer.

Shallow ConvNet [23]: Shallow ConvNet is specifically designed to decode band power features. It has two layers for temporal convolution and spatial filtering, respectively. These steps are similar to the bandpass filtering and CSP analysis of FBCSP.

EEGNet [24]: EEGNet starts with a temporal convolution to learn frequency filters and then uses a depthwise convolution to learn frequency-specific spatial filters. The separable convolution is a combination of a depthwise convolution, followed by a pointwise convolution, which learns how to optimally mix the feature maps together.

FBCNet [50]: FBCNet adopts depthwise convolution to extract spectral-spatial features from a multiview EEG representation, followed by a variance layer for feature extraction, which is the SOTA method for the BCI Competition IV 2a dataset.

MDRM [18]: Minimum distance to Riemannian mean (MDRM), is an implementation of the minimum distance to mean classification algorithm using Riemannian distance and Riemannian mean.

TSLDA [18]: Tangent Space Linear Discriminant Analysis (TSLDA) maps the EEG covariance matrix into the Riemannian tangent space to extract features, followed by dimension reduction using Principal Component Analysis (PCA) and Analysis of Variance (ANOVA). Linear Discriminant Analysis (LDA) is then employed as the classifier.

CSP-TSM [20]: CSP-TSM leverages both RTSM and theCSP algorithm for feature extraction, which are subsequently combined. Feature selection is performed using the Lasso regression analysis, followed by dimension reduction using LDA. SVM is chosen as the final classifier.

MTSMS [21]: Multiband Tangent Space Mapping with Subband Selection (MTSMS) first decomposes EEG signals into multiple subbands and estimates tangent features in each subband. Feature selection is carried out using an approach based on mutual information analysis. PCA and ANOVA are applied for feature dimension reduction, followed by classification using SVM.

FBRTS [22]: FBRTS employs a highly dense filter bank in combination with multiple time windows for EEG data processing. RTSM is used for feature extraction, and SVM serves as the classifier.

It's worth noting that, as described in Session 2.B, existing filtering strategies MDRM, TSLDA, and CSP-TSM all utilize the first filtering strategy, while MTSMS and FBCSP adopt the second filtering strategy. FBRTS employs the third filtering strategy.

C. Experimental Setups and Training Details

To assess the performance of DFBRTS, we conducted both within-session and cross-session evaluations. Following the methodology outlined in [33], within-session evaluation was performed using session 1 data from the OpenBMI dataset. For each subject, MI trials from both the training and test phases were utilized as training and test data, respectively. Cross-session evaluation, on the other hand, was conducted across entire sessions for both datasets. The training data came from session 1, while the test data were extracted from session 2. In line with previous studies [50], we employed a subset of 20 channels located in the motor region (FC-5/3/1/2/4/6, C-5/3/1/z/2/4/6, and CP-5/3/1/z/2/4/6) for preprocessing the OpenBMI dataset. It is important to note that our primary focus was on subject-specific MI decoding, and cross-subject validation was not the primary objective of this study. For both two datasets, time window of EEG data was selected begin with 0.5s after cue and end with ending of the trial. During DFB block, the level of DFB was set as 4 and filtering in each band is performed using a Chebyshev Type II filter.

Specifically, we applied regularization to covariance matrices ($M_i' = M_i + \alpha E$), where $E$ represents the unit matrix and $\alpha$ was set as $10^{-6}$, to address the issue of features containing covariance matrices with zero eigenvalues due to the influence of filtering. This regularization allowed for proper RTSM processing. The dropout probability for the CNN was set to 0.4, and the batch size during training was set to half of the training dataset size. Following the parameter settings from [47], the center loss was configured with λ=0.0005. For optimization, we employed the Adam optimizer with default settings (learning rate = 0.001, betas = 0.9, 0.999) [1-27] to minimize the loss function. In accordance with the study by [51], it has been observed that introducing an early-stopping mechanism by partitioning a validation set can introduce additional stochasticity to the algorithm's outcomes. Conversely, in practical applications, even in the absence of absolute certainty regarding the optimal model, it is feasible to achieve the desired level of decoding accuracy through ensemble methods, such as majority voting. Hence, we implemented a straightforward weight decay strategy, reducing the learning rate by a factor of 0.85 every 100 epochs, and set the maximum number of epochs to 2000.

## 5) Results

Table 2. Cross-Session classification accuracy compared with non-Riemannia-geometry-based methods on BCIC-IV-2a dataset

| Subject | FBCSP | EEGNet | Deep ConvNet | Shallow ConvNet | FBCNet | Proposed |
|---|---|---|---|---|---|---|
| 1 | 81.60 | 85.62 | 77.81 | 79.51 | 85.80 | **90.97** |
| 2 | 52.78 | 49.58 | 56.05 | 56.25 | 56.01 | **63.19** |
| 3 | 84.38 | **90.66** | 84.65 | 88.89 | 89.67 | 88.19 |
| 4 | 65.28 | 67.12 | 60.41 | **80.90** | 70.87 | 72.22 |
| 5 | 56.25 | 62.01 | 66.77 | 57.29 | 65.70 | **74.65** |
| 6 | 44.44 | 53.72 | **61.35** | 53.82 | 57.17 | 53.47 |
| 7 | 89.24 | 82.29 | 83.65 | 91.67 | 90.07 | **92.01** |
| 8 | 81.94 | 81.67 | 78.67 | 81.25 | 84.48 | **86.81** |
| 9 | 72.92 | **84.06** | 80.19 | 79.17 | 83.06 | 81.94 |
| Avg | 69.87** | 72.97* | 72.17* | 74.31 | 75.94 | **78.16** |
| Std | 15.90 | 15.13 | 10.64 | 14.54 | 13.73 | 12.59 |

The * and ** represent the statistically significant difference between the performance of DFBRTS and the given model with *: $p < 0.05$, and **: $p < 0.01$.

Table 3. Classification accuracy (Mean±Std) compared with non-Riemannia-geometry-based methods on OpenBMI dataset

| OpenBMI | FBCSP | EEGNet | Deep ConvNet | Shallow ConvNet | FBCNet | Proposed |
|---|---|---|---|---|---|---|
| Cross-session | 60.36± 14.84** | 69.45± 16.19* | 60.77± 11.31** | 61.30± 15.12** | 67.19± 14.24* | **71.58±** 15.81 |
| Within-session | 64.61± 16.92** | 70.89± 12.89 | 68.33± 12.23 | 67.08± 14.46* | 67.80± 13.47* | **72.97±** 16.13 |

Table 4. Cross-Session classification accuracy compared with Riemannia-geometry-based methods on BCIC-IV-2a dataset

| Subject | MDRM | TSLDA | CSP-TSM | MTSMS | FBRTS | Proposed |
|---|---|---|---|---|---|---|
| 1 | 84.03 | 84.72 | 81.60 | 85.42 | 85.42 | **90.97** |
| 2 | 58.33 | 51.74 | 53.12 | 51.74 | 57.29 | **63.19** |
| 3 | 71.53 | 87.85 | 81.94 | **88.89** | 85.07 | 88.19 |
| 4 | 64.93 | 48.26 | 62.50 | 63.19 | 63.19 | **72.22** |
| 5 | 44.44 | 39.58 | 40.97 | 56.94 | 57.64 | **74.65** |
| 6 | 45.49 | 50.35 | 49.31 | 51.39 | 49.31 | **53.47** |
| 7 | 59.60 | 82.29 | 72.57 | 82.64 | 82.99 | **92.01** |
| 8 | 72.57 | 85.07 | 83.68 | 79.17 | 83.68 | **86.81** |
| 9 | 77.43 | **83.33** | 76.74 | 82.29 | 80.56 | 81.94 |
| Avg | 63.93** | 68.13* | 66.94** | 71.30** | 71.68** | **78.16** |
| Std | 13.01 | 18.79 | 15.07 | 14.42 | 13.73 | 12.59 |

Table 5. Classification accuracy (Mean±Std) compared with Riemannia-geometry-based methods on OpenBMI dataset

| OpenBMI | MDRM | TSLDA | CSP-TSM | MTSMS | FBRTS | Proposed |
|---|---|---|---|---|---|---|
| Cross-session | 63.87±15.26** | 60.40±13.34** | 67.72±16.66** | 62.12±15.20** | 65.82±12.71** | **71.58±15.81** |
| Within-session | 66.16±15.09** | 62.70±16.74** | 71.09±16.34* | 56.87±8.92** | 66.47±13.82** | **72.97±16.13** |

  Table 2-4 presents a comparative analysis of the results between our proposed DFBRTS method and ten benchmark methods on two datasets. Overall, DFBRTS demonstrates superior classification accuracy. As seen from Tables 2 and 4, on dataset 2a, DFBRTS achieves an accuracy of 78.16% for the four-class MI decoding, significantly outperforming the ten benchmark methods. Tables 3 and 5 reveal that for the OpenBMI dataset, DFBRTS continues to exhibit notable improvements in average accuracy for binary classification tasks, achieving classification accuracies of 71.58% and 72.97% for cross-session and within-session evaluations, respectively. On some of the subjects, the DFBRTS method did not exhibit superior performance, potentially attributed to sample size influence and the non-uniform distribution of the covariance matrix, causing the data of these participants to potentially reside at the periphery of the manifold. Across all analyses in both datasets, DFBRTS consistently outperforms the ten benchmark methods by a significant margin. It is worth noting that, on the BCIC-IV-2a dataset, although DFBRTS has higher average accuracy and lower standard deviation compared to Shallow ConvNet and FBCNet, the significance test results are not statistically significant due to the smaller number of subjects. Except for the within-session analysis of CSP-TSM on the OpenBMI dataset, DFBRTS exhibits a highly significant performance difference compared to other Riemannian-geometry-based methods in all analyses. This underscores the effectiveness of our proposed methodin enhancing MI decoding performance.

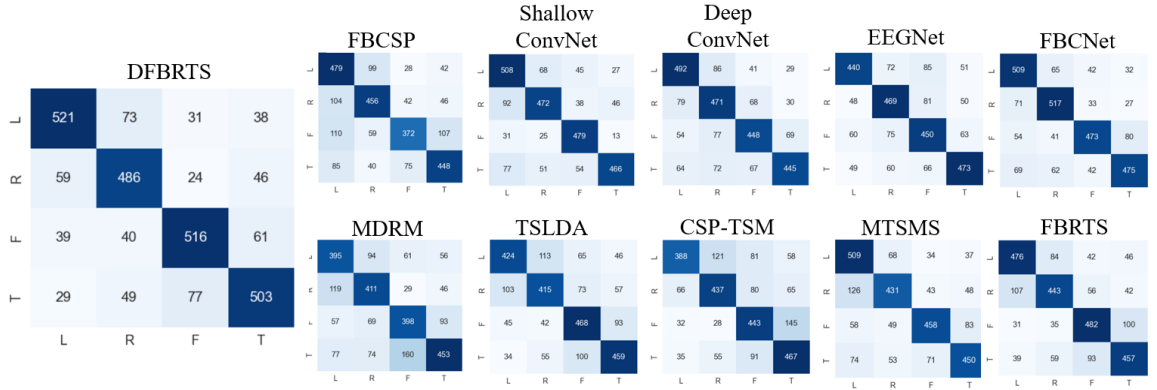

Fig. 3. Confusion matrices for DFBRTS and compared methods, where each column represents the actual values and each row depicts the predicted values of the model. The results are shown for the BCIC-IV-2a dataset, for cross-session analysis, where L, R, F, and T refer to MI of left hand, right hand, feet, and tongue, respectively.

Fig. 3 illustrates the confusion matrix results for the cross-session four-class MI tasks. Each method's confusion matrix results represent a composite of all subjects. As depicted in Fig. 3, DFBRTS exhibits the highest values along the diagonal and the smallest values off the diagonal for all datasets. Furthermore, FBCNet, which is the SOTA method for the BCIC-IV-2a dataset, appears to have more difficulty correctly identifying MI tasks other than those related to the right hand compared to DFBRTS. In addition, DFBRTS consistently demonstrates superior classification performance across various MI tasks.

Table 6. Decoding accuracy for subjects with the highest 25% and 25% lowest cross-session accuracy for the OpenBMI dataset

| Methods | FBCSP | Shallow ConvNet | Deep Convnet | EEGNet | FBCNet | MDRM | TSLDA | CSP-TSM | MTSMS | FBRTS | Proposed |
|---|---|---|---|---|---|---|---|---|---|---|---|
| Bottom 25% | 47.69± 1.93 | 46.38± 3.51 | 49.46± 2.27 | 51.38± 2.14 | 51.50± 1.83 | 49.73± 1.40 | 47.03± 4.03 | 51.31± 1.76 | 47.88± 1.85 | 52.69± 2.27 | **53.53± 4.00** |
| Top 25% | 83.04± 11.34 | 84.92± 9.28 | 77.19± 8.73 | 79.35± 7.32 | 88.30± 8.12 | 88.34± 7.90 | 80.08± 9.98 | 93.50± 5.06 | 86.46± 7.99 | 85.42± 6.62 | **94.84± 2.73** |

To conduct a more in-depth analysis of DFBRTS performance, we compared the accuracies of all methods for subjects falling within the top 25% and bottom 25% in terms of cross-session accuracy on the OpenBMI dataset. As presented in Table 6, among the top 25% of subjects, FBCSP achieves higher accuracy than Deep ConvNet and EEGNet but falls short of FBCNet, which aligns with the findings in [50], which shows that the design of the filter bank can improve the decoding ability of MI-EEG by acquiring the features of different frequency bands. It is well-known that deep learning methods are more prone to overfitting; therefore, on subjects exhibiting better performance and more pronounced features, Riemannian geometry-based methods demonstrate superior results. Moreover, FBCNet and CSP-TSM represent the best outcomes for deep learning and machine learning methods, respectively, with CSP-TSM exhibiting better performance. However, DFBRTS attains an even higher accuracy than both of them do. Turning to the bottom 25% of subjects, FBCNet performs the best among deep learning methods and surpasses all machine learning-based benchmark methods except for DFBRTS, on account of the improved feature extraction by integrating Riemannian geometry and CFC. Nevertheless, DFBRTS still achieves the highest average accuracy among the ten MI decoding methods.

## 6) Discussion

A. Feature Visualization Compared with Other Riemmania-geometry-based Methods

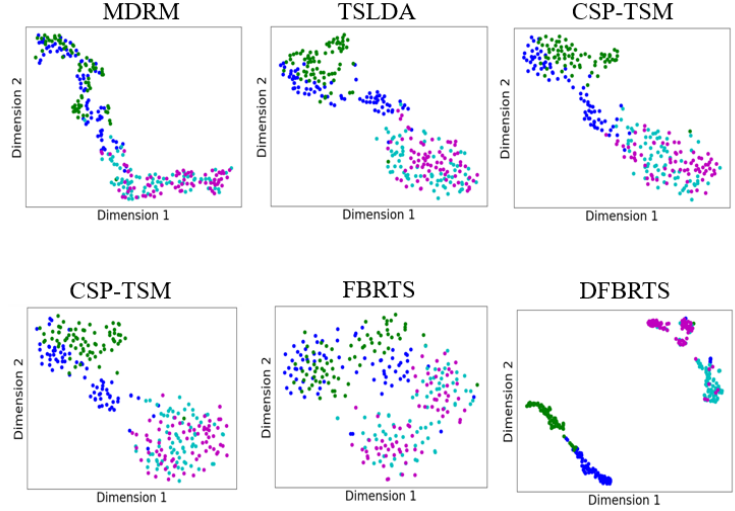

Fig. 4. Feature Visualization using two-dimensional t-SNE projection. The visualization is from subject A01 of BCI-IV-2a dataset. Features of DFBRTS are the features inputted into the classification layer, features of other methods are the features inputted into classifier. Specially, for MDRM, we use the distance from the target sample covariance matrix to various Riemannian mean points as features for t-SNE. Blue, green, cyan, purple points represent the MI of left hand, right hand, foot and tongue, respectively.

Fig. 4 displays feature visualizations of several Riemannian geometry-based methods, including the proposed method. It is evident that when the four motor imagery tasks are conceptually divided into "Left-Right" and "Feet-Tongue" groups, all Riemannian geometry-based methods exhibit strong discriminative capabilities for distinguishing between the "Left-Right" and "Feet-Tongue" categories. Furthermore, the five benchmark methods demonstrate good intra-category discrimination within the "Left-Right" group, while differentiation between Feet and Tongue features proves to be challenging. However, for DFBRTS, while it still enhances the distinctiveness of features between the "Left-Right" and "Feet-Tongue" groups, it significantly improves intra-group discriminative power, showcasing superior feature extraction capabilities.

B. Relation with DFB and CFC

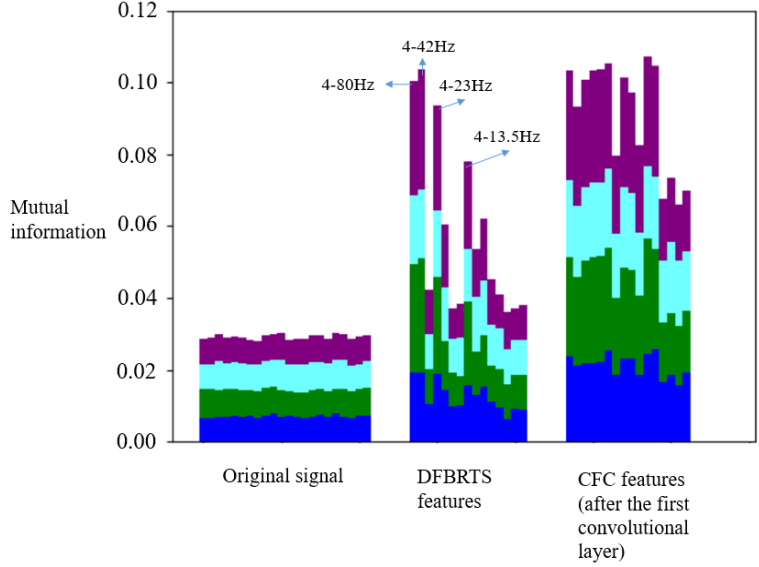

Fig. 5. Mutual information visualization of all channels in original signal, RTS features and CFC features. The visualization is from subject A01 of BCI-IV-2a dataset. Blue, green, cyan, purple points represent the MI of left hand, right hand, foot and tongue, respectively.

To illustrate the contribution of DFB's design and the incorporation of CFC concepts in the proposed method, we employed a visualization method based on mutual information theory. Mutual information is an information-theoretic approach that quantifies the information overlap between two sets of data, thereby indicating the effectiveness of features [52]. Figure 5 displays the mutual information between various channels and labels in the original signal, RTS features, and the first convolutional layer, which corresponds to the subband-wise convolutional layer. To provide a more intuitive representation of the influence of each channel on different classes, we performed additional processing when calculating mutual information. We set the selected class in the labels to 1 while assigning a value of 0 to all other classes, enabling the calculation of mutual information for the selected class. From Figure 5, it is evident that the mutual information in the original signal is relatively low. However, after feature extraction using DFBRTS, the mutual information values show a significant improvement.

Importantly, these frequency bands do not align with traditional complete rhythms such as Theta, Alpha, or Beta, nor do they correspond to fixed-frequency intervals within these rhythms. This underscores the effectiveness of DFB's design, as it allows for the extraction of RTS features in overlapped, multiscale frequency bands, which exhibit higher discriminative properties compared to features obtained through traditional filter bank designs. Furthermore, after passing through the first subband-wise convolutional layer, the CFC features, which are extracted by recombining RTS features from different frequency bands, contain additional information regarding frequency band interactions. These CFC features exhibit significantly higher mutual information values, indicating the effectiveness of the subband-wise convolutional layer in extracting CFC information.

C. Effects of DFB and DFB levels

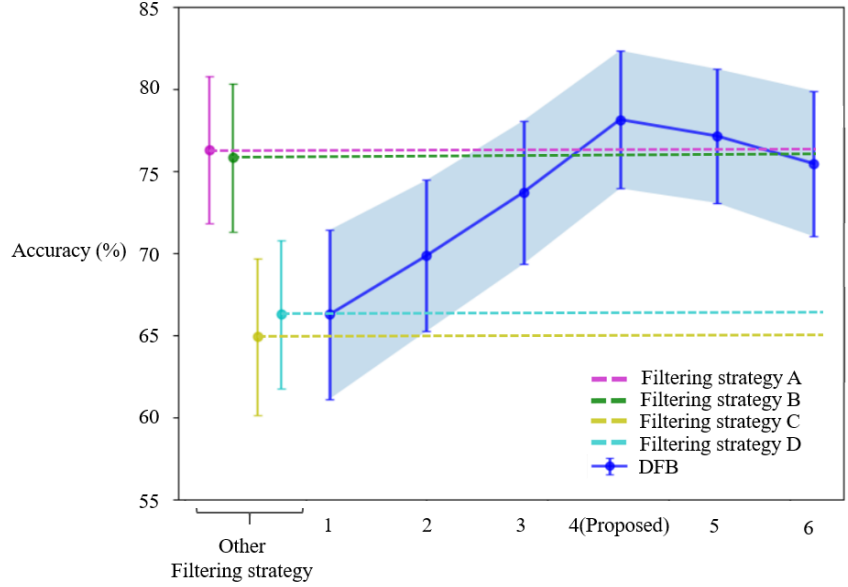

Fig. 6. Variation of decoding accuracy on BCIC-IV-2a dataset at different DFB levels (Mean± SEM). The dashed lines are the decoding accuracy of changing the DFB to another filtering strategy. Among them, filtering strategy A: non-overlapping filter bank with 4Hz as a band window in 4-30Hz; filtering strategy B: non-overlapping filter bank with 4Hz as a band window in 4-80Hz; filtering strategy C: 4-30Hz; filtering strategy D: 4-80Hz

Fig. 6 illustrates the variation in accuracy on the BCIC-IV-2a dataset as a function of the DFB level. It is evident that with only one decomposition, the method achieves results similar to early-stage Riemannian-geometry-based methods. This is because at this stage, the filters resemble traditional simple filters and cannot extract additional valuable information. Additionally, the CNN fails to obtain effective CFC features. As the decomposition level gradually increases, more valuable information becomes available, with the highest decoding accuracy achieved at level 4. However, with further increases in the decomposition level, the subbands in the DFB structure become overly narrow, leading to the generation of more suboptimal features. Consequently, there is a slight decrease in accuracy when the decomposition level reaches 5 and 6. Overall, the filter bank strategy (A and B) works better than a simple band window (C and D), but not as well as using DFB (proposed), which illustrates the effectiveness of DFB filtering strategy.

D.  Effects of CNN

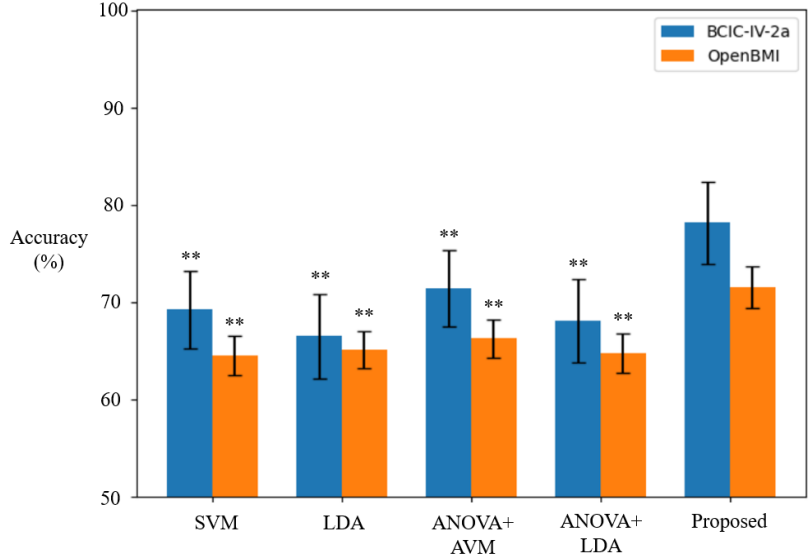

Fig. 7. Cross-session decoding accuracy with different classifier on both BCIC-IV-2a and OpenBMI dataset. Results are shown as the mean ± SEM (standard error of the mean (SEM)) of subjects. The * and ** represent the statistically significant difference between the performance of DFBRTS and the given model on the same dataset.

With all other settings kept constant, we replaced CNN with alternative machine learning methods to compare their accuracies, demonstrating the decoding capability of the proposed CNN-based approach. Gaussian kernel SVM and LDA, common classifiers, were employed for comparison. To further simulate traditional methods, we utilized PCA for orthogonal space mapping, followed by dimensionality reduction using ANOVA, serving as an additional comparison. The number of dimensions after ANOVA reduction was set to 50. Results from Fig. 7 reveal that SVM outperforms LDA. On the other hand, the inclusion of dimensionality reduction strategies further improved the accuracy of machine learning classifiers. Nevertheless, overall, the CNN-based classifier presented in this article outperforms all the compared machine learning classifier methods.

E. Effects of Center Loss

Table 6. Ablation of center loss in the cross-session analysis of OpenBMI dataset and BCIC-IV-2a dataset. Results are shown as the Mean ± Std of subjects.

|  | Subject | without center loss | with center loss (Proposed) |
| --- | --- | --- | --- |
| OpenBMI (Cross-session) | All | 70.72±14.37 | 71.58±15.81 |
|  | Acc. < 70% | 58.39±6.72* | 59.7±7.53 |
|  | Acc. > 70% | 87.64±9.26 | 88.35±10.95 |
| BCIC-IV-2a (Cross-session) | All | 77.62±12.41 | 78.16±12.59 |

In addition to the commonly used cross-entropy loss for classification tasks, we incorporated the center loss. The objective of the cross-entropy loss is to minimize misclassification of MI tasks. On the other hand, the center loss minimizes the sum of embedding space distances between sample features and their class feature centers, thereby making samples belonging to the same class more compact in the feature space [53]. To assess the impact of the center loss, we conducted ablation experiments on the center loss on both datasets. Specifically, we compared the performance of subjects with DFBRTS accuracy above and below 70% with and without the center loss. As shown in Table 6, for all 54 subjects in the OpenBMI dataset, the center loss resulted in a 0.86% increase in average accuracy. For subjects with accuracy below 70%, the center loss had a significantly positive impact on classification performance. However, for subjects with accuracy above 70%, the center loss had minimal impact on decoding performance. This validates that, as demonstrated in [47] and [54], the center loss can enhance the decoding performance for subjects with limited MI experience.

F. Influence of Activation Function

In this study, we utilized LeakyReLU instead of ReLU as the activation function. LeakyReLU allows input values below zero to pass through with a small negative slope, which helps alleviate the vanishing gradient problem and makes the model more sensitive to small gradient changes. We conducted cross-session analyses using both LeakyReLU and ReLU on BCIC-IV-2a and OpenBMI datasets separately. On both datasets, the performance of LeakyReLU as the activation function (OpenBMI: 71.58±15.81; BCIC-IV-2a: 78.16±12.59) exhibited slightly better results than ReLU (OpenBMI: 71.22±15.60; BCIC-IV-2a: 77.74±12.58), although the difference was not significant. The reason might be attributed to the fact that the Riemannian tangent space features we extracted are designed to represent data with non-Euclidean structures, and LeakyReLU's non-zero slope allows for greater flexibility to better adapt to the Riemannian tangent space feature data.

G. Limitations and Future Work

To clearly demonstrate the effectiveness of our proposed method, we chose standard fixed time windows and the same set of electrodes used in prior research (for the OpenBMI dataset). Furthermore, due to the limited training data in brain-computer interfaces (BCIs), recent studies have introduced data augmentation methods to further enhance the performance of deep neural networks [26, 55]. Additionally, given the scarcity of subject-specific data, many studies have employed adaptive transfer learning to fine-tune models for the target subjects [56]. In future work, we aim to combine time window selection and electrode selection methods to improve the capability of Riemannian geometry-based feature extraction. Data augmentation methods and transfer learning strategies will also be employed to enhance the decoding performance of DFBRTS.

# 7) Conclusion

In this paper, we introduce a MI-EEG decoding method called DFBRTS, which employs a dichotomous filter bank (DFB) and Riemannian tangent space mapping (RTSM) as the initial feature extraction steps, followed by a CNN comprising subband-wise and RTS-features-wise convolutional layers for further feature extraction and classification. To learn more discriminative features, we combine the center loss with the cross-entropy loss. Validation on two publicly available benchmark datasets underscores the exceptional performance of the DFBRTS method. DFBRTS surpasses other MI decoding algorithms with remarkable distinction on both datasets, achieving a notable classification accuracy of 78.16% for four-class and 71.58% for two-class hold-out classification. These results highlight the strong MI decoding capabilities of DFBRTS, highlighting its significant potential for applications in the field of MI-based BCI applications.